\documentstyle[preprint,pre,aps]{revtex}

\begin{document}

\draft
\title{Freezing Transition of Random Heteropolymers\\
Consisting of an Arbitrary Set of Monomers}
\author{Vijay S. Pande, Alexander Yu. Grosberg\dag, Toyoichi Tanaka}
\address{Department of Physics and Center for Materials Science and
Engineering, \\
Massachusetts Institute of Technology, Cambridge, Massachusetts
02139,
USA \\
\dag {\em On leave from:\/} Institute of Chemical Physics, \\
Russian Academy of Sciences, Moscow 117977, Russia}
\date{\today}

\maketitle
\begin{abstract}
Mean field replica theory is employed to analyze the freezing
transition of random heteropolymers comprised of an arbitrary number
($q$) of types of
monomers.  Our formalism assumes that interactions are short range
and heterogeneity comes only from pairwise interactions,
which are defined by an arbitrary $q \times q$
matrix.  We show that, in general,  there exists
a freezing transition from a random globule, in which the
thermodynamic equilibrium is comprised of an essentially
infinite number polymer conformations,
to a frozen globule, in which equilibrium ensemble is dominated by
one or very few conformations.
We also examine
some special cases of interaction matrices to analyze the
relationship between the freezing transition and the nature of
interactions involved.
\end{abstract}

\bigskip

\pacs{PACS 61.41.+h., 64.60.Cn, 87.15.Da, 64.60.Kw}

\section{Introduction}

The relationship between the sequence and conformation of a
heteropolymer
is one of the most challenging unsolved problems in biophysics.
In the case of proteins, it is widely believed that the native
functional
conformation is, in a sense, ``written'' in the sequence of the
heteropolymer in the ``language'' of the interactions between monomer
species.  This conformation is also believed to be both the ground
state from
thermodynamic point of view (better to say, it is structurally very
close to
the ground state, up to some short scale thermal and/or frozen
fluctuations)
and reliably accessible from the kinetic point of view.

The fact that even chains with random sequences can have a unique
frozen
ground state was first discussed in terms of phenomenological models
\cite{Wolynes}, where the freezing transition was shown to be
similar to that of the Random Energy Model (REM) \cite{Derrida}.
The REM-like freezing transition was also derived starting from
a microscopic Hamiltonian in which the interactions between pairs of
monomers were assumed to be random, independently taken from a
Gaussian
distribution \cite{IndepenInt}.
In this model, the nature of interactions between species was
parameterized in terms of the mean and width of the monomer-monomer
interaction distribution. Thus, in this sense, polymer sequence was
not explicitly included in this model, since it is absent from the
Hamiltonian.
As for models with polymer sequences explicitly present, two have
been considered so far:
2 letter Ising-type model \cite{Sfatos} and
the so-called $p$-charge model \cite{OrlandP,ShakP}. These models
were shown to
also exhibit a freezing phase transition for random chains.

Therefore, it is natural to conjecture that any sort of
random heteropolymer will have this kind of transition,
and the question is whether we are able to understand
the properties and characteristic temperature of this
transition for realistic models of heteropolymers.
Indeed, proteins, for example, are comprised of 20
kinds of different monomers, which interact to each
other in a complicated manner. There are several
relevant types of  interactions between different
monomers, such as van-der-Waals interactions,
dipole-dipole interactions, hydrogen bonds, and
hydrophobic interactions.

As long as we are speaking about short-range interactions,
interactions can be described in terms of a matrix: if there are $q$
types of monomers, we have a $q \times q$ matrix, where each $(i,j)$
matrix element represents the energy of interaction between monomers
of the types $i$ and $j$, given that they are in spatial contact.
There were several attempts in the literature to derive this kind of
``interaction'' matrix for real amino acids (see, in particular,
\cite{Jernigan}). It is rather difficult, however, to derive this
kind of matrix.  Furthermore, the sensitivity of heteropolymer
properties to deviations of the interaction matrix is unclear. For
computer simulations, for example, it is important to know how
precise one should be in choosing the interaction energies in order
to reproduce the native state and to avoid the appearance of some
other state, structurally completely different, which may appear as
the ground state of a simulated system due to an imperfect
interaction matrix. Of course, other non-protein heteropolymers might
be also of interest.

In this paper, we consider the freezing transition for a
heteropolymer
with an arbitrary interaction matrix. We derive a general formalism
for the
analysis of the freezing transition of random chains in which only
short range interactions are assumed. In addition to the formal
benefit that the general treatment establishes a formalism with which
other short range species interaction models can be derived as
special cases by using specific interaction matrices, this theory can
be used to analyze what properties of a species-species interactions
matrix effect the freezing transition and in what way.

\section{Development of the Formalism}

\subsection{The Model and its Hamiltonian}

Consider a heteropolymer chain with a frozen sequence of monomers
$s_I$, where
$I$ is the number of monomer along the chain ($1 \leq I \leq N$) and
$s_I$ is
the sort of monomer $I$ in the given sequence. Let $q$ be the total
number of
different monomer species, $1 \leq s(I) \leq q$.
In the condensed globular state, the spatial structure of the chain
is governed
by volume interactions between monomers. The disorder
and heteropolymer effects of different monomer species comes mainly
through pairwise monomer-to-monomer interactions. On the other
hand, higher order interactions provide the non-specific excluded
volume effect, while chain connectivity defines the set of available
placements of monomers in space. This is clear when one considers
the lattice model, where subsequent monomers are nearest neighbors on
the
lattice (chain connectivity): a site on the lattice can be occupied
by only one
monomer (excluded volume effect), and the energy is given as a sum of
pairwise
interactions of the nearest neighbors on the lattice. The complicated
set
of monomer-monomer interactions, related to frozen-in sequence,
appears then due to the restricted set of pairings of monomers in the
space. The interaction part of Hamiltonian
can be therefore written in a rather simple way:
\begin{equation}
{\cal H} =  \sum_{i,j}^q \sum_{I,J}^N B_{ij} \delta({\bf r}_I - {\bf
r}_J)
\delta (s_I,i) \delta (s_J,j) + {\cal H}'
\label{eq:GenHam}
\end{equation}
where $B_{ij} \delta({\bf r}_I - {\bf r}_J)$ gives the Mayer function
of
short range interaction between monomers of {\em species\/} $i$ and
$j$, placed
in space at the distance ${\bf r}_I - {\bf r}_J$ apart from each
other,
$s_I$ is the species of monomer number $I$ (``spin'' of monomer $I$),
and
$\delta$ is either Kroneker or Dirac delta. Eq~(\ref{eq:GenHam}) has
the simple interpretation that monomers number $I$ and $J$ interact
based upon
their
proximity, $\delta ({\bf r}_I - {\bf r}_J)$, and the second virial
coefficient of interaction between the species of the two monomers,
$B_{s_I
s_J}$.  The ${\cal H}'$ contribution contains all higher order
interactions
of monomers.  We assume that it is ``homopolymeric'' in form, i.e. it
does not
depend on the monomer species, but only on the overall density
$\rho$.  It can be written as ${\cal H}' = C \rho^2 + D \rho^3 +
\ldots$, where all virial coefficients $C, D, \ldots$ are assumed to
be positive (repulsive).

Throughout the paper, we will use the following notation:
upper case Roman characters label monomer numbers, i.e. bead number
along the
chain ($1 \leq I \leq N$),
lower case Roman characters label monomer species numbers ($1 \leq i
\leq q$),
and lower case Greek characters are for replica indices ($1 \leq
\alpha \leq
n$),
which will be defined later. We will be also using the notation for
vectors and
operators (matrices) with the clear indication of the dimensionality
of the
corresponding space, as we consider several different spaces
simultaneously.
For example, the interaction matrix with matrix elements $B_{ij}$
will be
denoted as $\widehat{B}^{(q)}$. In this notation, vector
$\vec{\rho}^{(q \infty )}$
means the density distribution $\rho_i ({\bf R})$ for all species
($q$) over
$3D$ ${\bf R}$-space ($\infty$).

\subsection{Replicas}

The statistical mechanics of a heteropolymer chain is expressed
through the partition function, which can be somewhat formally
written as
\begin{equation}
Z({\rm seq})=\sum_{\rm conformations} \exp \left[- {1 \over T} {\cal
H} \left(
{\rm conf, seq} \right) \right] \ ,
\label{symbZ}
\end{equation}
where we have clearly indicated that our Hamiltonian depends on both
conformation and sequence. The standard way to approach the partition
function of a system with frozen disorder is to employ, first, the
principle
of self-averaging of free energy and, second, the replica trick:
\begin{equation}
F = \langle F({\rm seq}) \rangle_{\rm seq} = -T\langle \ln Z({\rm
seq})
\rangle_{\rm seq} = \lim_{n \rightarrow 0} {  \langle Z^n({\rm seq})
\rangle_{\rm seq} -1  \over n} \ ,
\label{rep}
\end{equation}
where $\langle \ldots \rangle_{\rm seq}$ means average over the set
of all
possible $q^N$ sequences.

In this paper, we consider random sequences, meaning that the species
$1,2,
\ldots ,q$ appear independently along the chain with the
probabilities $p_1,
p_2, \ldots, p_q$ ($\left\{ p_i \right\} ={\vec p}^{(q)}$), so that
the
probability of realization of the given sequence $({\rm seq} = s_1,
s_2,
\ldots, s_I, \ldots , s_N)$ is written as
\begin{equation}
P_{\rm seq} = p_{s_1}   p_{s_2}  \ldots   p_{s_I}  \ldots  p_{s_N} =
\prod_{I=1}^N p_{s_I}
\end{equation}

Collecting the above equations, we can write the key value of
$n$-replica
partition function as
\begin{eqnarray}
\langle Z^n({\rm seq}) \rangle_{\rm seq} & = &  \sum_{\rm seq} P_{\rm
seq}
\left\{ \sum_{\rm conformations} \exp \left[- {1 \over T} {\cal H}
\left( {\rm
conf, seq} \right) \right] \right\}^n    \nonumber \\
& = &  \sum_{\rm seq} P_{\rm seq} \sum_{ C_1, \ldots , C_n} \exp
\left[ - { 1
\over T} \sum_{\alpha = 1}^{n} {\cal H} \left( C_{\alpha}, {\rm seq}
\right)
\right]   \ ,
\end{eqnarray}
where $C_{\alpha} =C_1, \ldots , C_n$ stand for conformations of
replica number
$\alpha$.

For each conformation and each replica, we introduce density
distributions of
all species as
\begin{equation}
m_i^{\alpha}({\bf R}) =\sum_{I=1}^N \delta (s_I,i) \delta({\bf
r}_I^{\alpha} - {\bf R}) \ ; \ \ \left\{ m_i^{\alpha}({\bf R})
\right\}
\equiv {\vec m}^{(qn \infty)}.
\end{equation}
For simplicity, we will not explicitly include the sequence
independent terms ${\cal H}'$ from the original
Hamiltonian~(\ref{eq:GenHam}).
We then write in terms of the densities
\begin{eqnarray}
{\left< Z^n({\rm seq}) \right>}_{\rm seq} & = & \sum_{\rm seq} P_{\rm
seq}
\sum_{ C_1, \ldots , C_n}
 \exp \left\{ - {1 \over  T } \sum_{\alpha = 1}^n  \sum_{i,j=1}^q
\int \!
d{\bf R}_1 d{\bf R}_2 \ m_i^{\alpha}({\bf R}_1) B_{ij} \delta({\bf
R}_1
- {\bf R}_2) m_j^{\alpha}({\bf R}_2) \right\} \nonumber \\ & = &
\sum_{\rm
seq} P_{\rm seq} \sum_{ C_1, \ldots , C_n}
 \exp \left\{ - {1 \over  T }
\left< {\vec m} \left| {\widehat B} \right| {\vec m} \right>^{(qn
\infty)}
\right\} \ ,
\end{eqnarray}
where $\left< \left| \ldots \right| \right>^{(qn \infty)}$ means
scalar product
in which all vectors and operators are supposed to have
dimensionality as
indicated ( $q \times n \times \infty$ in this case). Operator
${\widehat B}^{(qn
\infty)}$ is $B_{ij}$ with respect to monomer species, and it is
diagonal in
both replica space and real coordinate space, meaning that it has
matrix
elements $B_{ij} \delta_{\alpha \beta} \delta ({\bf R}_1 - {\bf
R}_2)$. The
next step is to perform Hubbard-Stratonovich transformation
\cite{Sfatos} of the form
\begin{equation}
\langle Z^n({\rm seq}) \rangle_{\rm seq} =  {\cal N} \sum_{C_1,
\ldots , C_n}
\int \! {\cal D}\! \left\{ \phi \right\} \  \exp \left\{ {T \over 4}
\left< {\vec \phi} \left| {\widehat B}^{-1} \right| {\vec \phi}
\right>^{(qn
\infty)} \right\}
\times \sum_{\rm seq} P_{\rm seq} \exp \left\{
\left< {\vec \phi} \left|  \right. {\vec m} \right>^{(qn \infty)}
\right\} \ .
\end{equation}
Here $\left\{ \phi _i^{\alpha} ({\bf R}) \right\} = {\vec \phi}^{(qn
\infty)}$
are the fields conjugated to the corresponding densities and ${\cal
N}$ is
normalization factor which comes from integration over $\phi$.

Note that the sum over sequences enters only in the last ``source''
term of
this expression:
\begin{equation}
\exp \left\{ {\rm source \ term} \right\}  =\sum_{\rm seq} P_{\rm
seq} \exp
\left\{
\left< {\vec \phi} \left|  \right. {\vec m} \right>^{(qn \infty)}
\right\} \ .
\end{equation}
The summation, or average, over the sequences is easier to describe
in non-vector notation:
\begin{eqnarray}
\exp \left\{ {\rm source \ term} \right\} & = & \sum^q_{s_1,s_2,
\ldots , s_N }
\prod_{I=1}^N  p_{s_I} \prod_{i=1}^q \exp \left\{ \delta (s_I , i )
\sum_{\alpha = 1}^n \int \! d{\bf R} \  \phi_i^{\alpha}({\bf R})
\delta (
{\bf r}_I^{\alpha} - {\bf R}) \right\} \nonumber \\
& = & \prod_{I=1}^N \sum_{s_I=1}^q p_{s_I} \prod_{i=1}^q \exp \left\{
\delta
(s_I , i )  \sum_{\alpha = 1}^n \int \! d{\bf R} \
\phi_i^{\alpha}({\bf R})
\delta ( {\bf r}_I^{\alpha} - {\bf R})  \right\} \nonumber \\
& = & \prod_{I=1}^N \sum_{i=1}^q p_i \exp \left\{  \sum_{\alpha =
1}^n \int \!
d{\bf R} \  \phi_i^{\alpha}({\bf R}) \delta ( {\bf r}_I^{\alpha} -
{\bf R})
\right\}
\end{eqnarray}

As in case of two-letter heteropolymer, to extract the relevant order
parameters, we expand over the powers of the fields $\phi$ (high
temperature
expansion) and keep terms up to ${\cal O}(\phi^2)$:
\begin{eqnarray}
{\rm source \ term} & = & \sum_{i=1}^q \sum_{\alpha=1}^n \int \!
d{\bf R} \rho
^{\alpha} ({\bf R}) p_i \phi ^{\alpha}_i ({\bf R}) \nonumber \\ & + &
{1
\over 2} \sum_{i,j=1}^q \left[ p_i \delta _{ij} - p_i p_j \right]
\sum _{\alpha
, \beta =1} ^n \int \! d{\bf R}_1 \! \int \! d{\bf R}_2 \;
\phi_i^{\alpha}({\bf R}_1) Q_{\alpha \beta} ( {\bf R}_1, {\bf R}_2)
\phi_j^{\beta}({\bf R}_2) \ ,
\label{eq:source_term_series}
\end{eqnarray}
where we use standard definitions \cite{Orland,IndepenInt,q2Design}
\begin{equation}
Q_{\alpha_1 , \ldots , \alpha_k } ({\bf R}_1, \ldots, {\bf R}_k)
= \sum_{I=1}^N \prod_{\kappa = 1}^k \delta({\bf
r}_I^{\alpha_{\kappa}} - {\bf
R}_{\kappa}) \ ,
\end{equation}
\begin{equation}
Q_{\alpha} ({\bf R}) \equiv \rho ^{\alpha} ({\bf R})= \sum _{I=1}^N
\delta
({\bf r}_I^{\alpha} - {\bf R} ) \ \ \ ; \ \ \ Q_{\alpha \beta} ({\bf
R}_1,
{\bf R}_2) = \sum _{I=1}^N \delta ({\bf r}_I^{\alpha} - {\bf R}_1 )
\delta
({\bf r}_I^{\beta} - {\bf R}_2 ) \ .
\end{equation}
Note that the total density of the polymer chain $\rho ^{\alpha}
({\bf R})$
in equilibrium does not depend on replica number  and, within a large
globule,
does not depend on ${\bf R}$.  Replicas are  interpreted as pure
states of
the polymer chain \cite{Mezard,IndepenInt,q2Design}, and the
$k$-replica
order parameter $Q_{\alpha_1, \ldots , \alpha_k}$ is interpreted as
the overlap
between
replicas $\alpha_1, \ldots, \alpha_k$.

The $n$-replica partition function is now written in the form:
\begin{eqnarray}
&&\langle Z^n({\rm seq}) \rangle_{\rm seq}  =  {\cal N} \sum_{C_1,
\ldots ,
C_n}\int \! {\cal D}\! \left\{ \phi \right\} \nonumber \\ & \times &
\exp \left\{ {
\left< {\vec \phi} \right| {T \over 4} B_{ij}^{-1} \delta ({\bf R}_1
- {\bf R}_2)
\delta_{\alpha \beta} +
{1 \over 2} Q_{\alpha \beta}({\bf R}_1,{\bf R}_2)  \Delta_{ij} \left|
{\vec \phi} \right>
}^{(q n \infty )} +
\left< {\vec \rho}  \left| \right.  {\vec \phi} \right>^{(n q
\infty)}
\right\}
\label{eq:GaussInt0}
\end{eqnarray}
where
\begin{equation}
\Delta_{ij} = p_i \delta_{ij} - p_i p_j \ \ {\rm and} \ \ \
{\vec \rho}^{(q n \infty)}  \equiv \rho_i^\alpha({\bf R}) =
p_i \sum_{I=1}^N \delta({\bf r}_I^\alpha - {\bf R}) \ .
\end{equation}

We are left with a Gaussian integral~(\ref{eq:GaussInt0}) for the
$n$-replica
partition function, which
is simplified by the argument given in
\cite{IndepenInt,Sfatos,q2Design},
showing that the ${\bf R}$-dependence of $Q_{\alpha \beta}$ is of
$\delta$-type, so that
\begin{equation}
Q_{\alpha \beta}({\bf R}_1,{\bf R}_2) =\rho q_{\alpha \beta}\delta
({\bf
R}_1 - {\bf R}_2) \ \
\ ,
\end{equation}
where diagonal matrix elements of new matrix ${\widehat q}^{(n)}$ are
1, while off-diagonal elements are either 0
or 1. This means physically that two replicas $\alpha$ and $\beta$
might be
either uncorrelated (independent), so that $Q_{\alpha \beta}=0$, or
they may be
correlated so that one repeats the $3D$ fold of the other down to the
microscopic length scale, so that $Q_{\alpha \beta}({\bf R}_1,{\bf
R}_2)
=\rho \delta ({\bf R}_1 - {\bf R}_2)$.
We do not repeat this argument here, as it is explained elsewhere
(see the
argument presented in \cite{q2Design} which is slightly different
from the
original one \cite{IndepenInt}).

\subsection{Effective Energy in Replica Space}

With the simplified form of $Q$ matrix, we evaluate the Gaussian
integral over
all $\phi^\alpha$ variables. This yields
\begin{equation}
\langle Z^n({\rm seq}) \rangle_{\rm seq} =   \sum_{C_1, \ldots , C_n}
\
\exp \left[ - N  E \left\{Q \right\}
\right]
\end{equation}
with the energy of the form
\begin{eqnarray}
E & = &
 \left< {\vec \rho}^{(n q)} \left| \left[ T \left( {\widehat B}^{(q)}
\right)^{-1} \otimes {\widehat I}^{(n)} + 2 \rho
{\widehat q}^{(n)} \otimes {\widehat \Delta}^{(q)} \right]^{-1}
\right| {\vec \rho}^{(n q)}
\right>
 \nonumber \\
&& +
{1 \over 2} \ln \det \left[ {T \over 4} \left( {\widehat B}^{(q)}
\right)^{-1} \otimes {\widehat I}^{(n)} +
{1 \over 2} \rho {\widehat q}^{(n)} \otimes {\widehat \Delta}^{(q)}
\right] +
{1 \over 2} \ln \det \left(4 {\widehat B}^{(qn)} /T \right)
	   \ .
\label{eq:lndet}
\end{eqnarray}
Here $\otimes$ means the direct product, eg. for the block matrix
${\widehat B}^{(qn\infty)} = B_{ij} \delta_{\alpha \beta} \delta
({\bf R}_1 - {\bf R}_2)
 = {\widehat B}^{(q)} \otimes {\widehat I}^{(n)} \otimes {\widehat
I}^{(\infty)}$. In general, ${\widehat A}^{(r)} \otimes {\widehat
B}^{(s)}$ produces block matrix of the total size $rs$, according to
the rule: instead of each matrix element of  ${\widehat A}^{(r)}$
matrix, say $A_{uv}$, we substitute  the block equal to $A_{uv}
{\widehat B}^{(s)}$. The last term in~(\ref{eq:lndet}) comes from
normalization factor ${\cal N}$ in
eq~(\ref{eq:GaussInt0}); it is easy to check that the normalization
factor created by Gaussian integration ${\cal N}$, simply eliminates
normalization factors first introduced by the Hubbard-Stratonovich
transformation.
%
%
%
%
Noting that $\det({\widehat A}) \cdot \det({\widehat B}) =
\det({\widehat A}{\widehat B})$, we can simplify the last
relationship as
\begin{eqnarray}
E & =  & {1 \over 2}  \ln \det \left[ \widehat{I}^{(qn)}+{2 \rho
\over T}
 {\widehat q}^{(n)} \otimes {\widehat \Delta}^{(q)}
{\widehat B}^{(q)} \right] + \nonumber \\
 & + & \left< {\vec \rho}^{(n q)} \left| {1 \over T} {\widehat
B}^{(q)} \otimes {\widehat I}^{(n)}
\left[ \widehat{I}^{(qn)}+{ 2 \rho \over T}
{\widehat q}^{(n)} \otimes {\widehat \Delta}^{(q)} {\widehat B}^{(q)}
 \right]^{-1} \right| {\vec \rho}^{(n q)} \right> \ ,
\label{eq:energy1}
\end{eqnarray}

We have assumed that the Gaussian integral converges and can be
calculated. This is guaranteed only by the appropriate form of
${\widehat q}^{(n)}$ matrix, i.e., by replica symmetry breaking. We
make an {\em ansatz\/} that ${\widehat q}^{(n q)}$
is of the form of a Parisi matrix with one-step replica symmetry
breaking
\cite{Parisi,Sfatos}.
We say that replicas can be gathered into $n/x$ groups each of which
consists
of $x$ replicas.
The conformations of all of the replicas in a given group coincide to
the
microscopic scale, i.e. for $\alpha, \beta \in$ group A and $\gamma
\in$ group
B, then $q_{\alpha \beta}=1$
and $q_{\alpha \gamma}=q_{\beta \gamma}=0$.
Thus ${\widehat q}^{(n q)}$ can be written as a block matrix (in
replica space) which
is partitioned into $n/x$ blocks of size $x \times x$ along the
diagonal.   Inside each diagonal block, $q_{\alpha \beta}=1$, and
outside
$q_{\alpha \beta}=0$.
In fact,
it was recently shown that this form can be derived by energy
minimization in
the two letter case \cite{q2Design}, and we can easily repeat this
argument for
the general $q$-letter case at
hand. For the sake of simplicity, however, we omit the derivation,
thus,
formally employ the ansatz.

We can substantially simplify both terms in the
energy~(\ref{eq:energy1}), and convert them into the form
\begin{equation}
E = {n \over 2x} \ln  \det \left({\widehat I} + {2 \rho x \over T}
{\widehat \Delta} {\widehat B}
\right) +
n \left< \vec{p} \left| \rho \  { {\widehat B} \over T} {\left(
{\widehat I}  +
{2 \rho x \over T} {\widehat \Delta} {\widehat B}  \right)}^{-1}
\right| \vec{p} \right>
\label{eq:energy}
\end{equation}
Here we have dropped the labels of the dimensionality of the vectors
and
operators, as all of them are of the same dimensionality $(q)$. This
is because we have diagonalized the energy in both $\bf R$ ($\infty$)
and
replica ($n$) spaces, so only the species dimension ($q$) remains.

The proof of the simplification leading to~(\ref{eq:energy}) is given
in the Appendix. We now turn to its analysis.

\subsection{Effective Entropy in Replica Space}

%
%
In order to get the free energy, we must also consider the entropy
change
due to the constraint on $q_{\alpha \beta}$.  Following
Refs.~\cite{Sfatos,Grosberg},
we argue that due to the polymeric bonds connecting monomers along
the
chain, once one monomer is fixed in space, the next must be placed
within a volume
$a^3$, where $a$ is the distance between monomers along the chain.
Since replicas that belong in the same group coincide within a tube
of radius $R_t \sim v^{1 \over 3}$, where $v$ is the excluded volume
of a single monomer, there are $a^3/v$ ways to place the next monomer
and thus the entropy per monomer in just $\ln(a^3/v)$. But since all
replica conformations coincide within the group, we must restrict the
position of the next monomer to a single place.  Following the Parisi
ansatz for one-step {\sc rsb}, for $n$ replicas, there are $n/x$
groups with $x$ replicas per group.  The entropy loss is therefore
\begin{equation}
S = N s { n \over x} (x-1)
\label{eq:entropy}
\end{equation}
where $s=\ln(a^3/v)$ is related to the flexibility of the chain.

\subsection{Freezing Transition}

Recall that, for notational convenience, we drop the indication of
dimensionality, as all operators and vectors are now assumed to be in
species space, i.e. dimensionality $q$. We optimize the free energy
\begin{equation}
F ={n \over 2x} \ln  \det \left({\widehat I} + {2 \rho x \over T}
{\widehat
\Delta} {\widehat B}\right)
+ n \left<  \vec{p} \left| {\rho \over T} {\widehat B} { \left(
\widehat{I} + {2 \rho x
\over T} \widehat{\Delta} {\widehat B} \right) }^{-1} \right| \vec{p}
\right> + s { n
\over x} (x-1)
\label{eq:FreeEnergy}
\end{equation}
with respect to $x$, yielding
\begin{eqnarray}
2 s & = & \ln  \det \left({\widehat I} + {2 \rho x \over T} {\widehat
\Delta} {\widehat B}
\right)  -
{\rm Tr} \left[ {2 \rho x \over T} {\widehat \Delta} {\widehat B}
{\left({\widehat I} +
{2 \rho x \over T} {\widehat \Delta} {\widehat B} \right) }^{-1}
\right] \nonumber \\
& & +
 \left<  \vec{p} \left| {2 \rho x \over T} {\widehat B} {2 \rho x
\over T} {\widehat \Delta}
{\widehat B} { \left( \widehat{I} + {2 \rho x \over T}
\widehat{\Delta} {\widehat B} \right)
}^{-2} \right| \vec{p} \right> \ .
\end{eqnarray}

As is clear from the very structure of this equation, its solution is
of the form $x = T \xi/2 \rho$, where $\xi$ is given by
\begin{equation}
2s = \ln  \det \left({\widehat I} + \xi {\widehat \Delta} {\widehat
B} \right)  -
{\rm Tr} \left[ \xi {\widehat \Delta} {\widehat B}  {\left({\widehat
I} + \xi {\widehat \Delta}
{\widehat B} \right) }^{-1} \right] +
\left<  \vec{p} \left| \xi^2 {\widehat B} {\widehat \Delta} {\widehat
B}  { \left( \widehat{I}
+  \xi {\widehat \Delta} {\widehat B} \right) }^{-2} \right| \vec{p}
\right> \ .
\label{eq:finalselfconsist}
\end{equation}
Recall that $x$ is the number of replicas in one group, i.e., the
number of replicas which have the same conformation down to
microscopic fluctuations. This interpretation is clear when $n$ is
integer and $n>1$. While taking the $n \rightarrow 0$ limit, we have
to consider $x$ to be in between $n$ and $1$, so that $x<1$ means the
existence of grouping of replicas, or broken replica permutation
symmetry, while $x$ approaching 1 means the restoration of replica
symmetry. Therefore, $x=1$ defines the point of phase transition
between the frozen globular phase with broken replica symmetry and
the phase of a random ``liquid-like'' replica symmetric globule. The
corresponding freezing temperature is given by $T_f=2 \rho/\xi$.

Thus, from the $n$-replica free energy, we obtain the real free
energy

\begin{equation}
F =
\cases{
 {T_f \over 2} \ln  \det \left({\widehat I} + {2 \rho \over T_f}
{\widehat
\Delta} {\widehat B}\right)
+ \left<  \vec{p} \left| \rho {\widehat B} { \left( \widehat{I} + {2
\rho \over T_f}
 \widehat{\Delta} {\widehat B} \right) }^{-1} \right| \vec{p} \right>
- s  \left(T_f - T \right)
& for $T<T_f$ \cr
 {T \over 2} \ln  \det \left({\widehat I} + {2 \rho \over T}
{\widehat
\Delta} {\widehat B}\right)
+ \left<  \vec{p} \left| \rho {\widehat B} { \left( \widehat{I} + {2
\rho \over T}
\widehat{\Delta} {\widehat B} \right) }^{-1} \right| \vec{p} \right>
& for $T>T_f$ \cr
}
\end{equation}

%
%

\section{Discussion}

\subsection{What is ${\widehat \Delta}$?}

We first examine the physical meaning of the operator ${\widehat
\Delta}$ and the term ${\widehat \Delta} {\widehat B}$.  From
the
definition of ${\widehat \Delta}$, we have
\begin{equation}
{\left( {\widehat \Delta} {\widehat B} \right) }_{ik} = \sum_j
\left(
p_i \delta_{ij} - p_i p_j \right) B_{jk} = p_i B_{ik} - \sum_j  p_i
p_j B_{ij}
\label{eq:DB}
\end{equation}
We can always write $B_{ij}$ in terms of a sum of a homopolymeric
attraction ($B_0$) and heteropolymeric deviations
 ($b_{ij}=B_{ij}- \left< B  \right>$).
{}From~(\ref{eq:DB}), we see that ${\widehat \Delta}$ removes the
mean interaction of species $k$ from all matrix elements $B_{kj}$.
In other words, ${\widehat \Delta}$ removes all homopolymeric
effects.

It is instructive to examine what happens to the
energy~(\ref{eq:energy}) when one formally takes $ {\widehat \Delta}
{\widehat B} = 0$; in this case
\begin{equation}
E=n \left( \rho / T \right) \left< {\vec p} \left|   {\widehat B}
\right|  {\vec p}  \right>  = n \left( \rho / T \right) \sum_{ij} p_i
p_j  B_{ij} = n  \left( \rho / T \right) \left< B  \right>  \ ,
\label{eq:secvirterm}
\end{equation}
which is simply the averaged second virial term. Note that as this
term is not coupled to $x$, $ \left< B  \right>$ does not
enter into the calculations of the freezing temperature.
We note that the terms $n {\cal H}' = n C \rho / T + \ldots$ from the
original
Hamiltonian~(\ref{eq:GenHam}) are not explicitly written, but must be
considered when
optimizing the free energy.
Thus, for $| \left< B  \right> | \gg | b_{ij}|$,
we can optimize the free energy with respect to $x$ and $\rho$
independently.
However, if this condition is not valid, the coupling between density
and the replica overlap order parameter becomes significant; this
should lead to other interesting physical phenomena, which are beyond
the scope of this paper.
The ``homopolymeric'' attractive second virial term, in competition
with the repulsive higher order terms in ${\cal H}'$, is responsible
for the formation and maintenance of the globular conformation with a
reasonably high density.  Therefore, $\left< B  \right>$ primarily
enters into homopolymer effects, such as the coil to globule
transition. Other effects, such as the freezing
transition, are purely heteropolymeric, and are due to $b_{ij}$, or
${\widehat \Delta} {\widehat B}$ terms; they are related to the
choice of some energetically preferential conformations out of the
total vast number of globular conformations.

For the homopolymer case ($q=1$ or $B_{ij}= B_0$) or
the effective homopolymer case (a heteropolymeric
interaction matrix is rendered homopolymeric due to the choice of
composition $\vec{p}$; say, $p_1=1$, while others $p_i=0$), we
immediately see that ${\widehat \Delta} {\widehat B} =
0$, so $T_f=0$ and thus there is no
freezing transition. (This is of course just trivial check of
consistency of our equations).


%
%

\bigskip

\subsection{Two Exactly Solvable Models}

There are some models which can be solved exactly from
eq~(\ref{eq:finalselfconsist}).
We will see that the exact solution of simple models yields insight
which will be
important in the more general consideration of the next section.

\subsubsection {Potts Model}

 Potts interactions are defined by the interaction
matrix $B_{ij} = b \delta_{ij} + B_0$.
The freezing temperature can be found exactly for this model
for the case of even composition, i.e. $p_i = 1/q$.
{}From (\ref{eq:energy}), we see that the relevant matrix to address
is
${\widehat I} + 2 \rho x {\widehat \Delta} {\widehat B}/T$.
As the diagonal elements of this matrix are
$1 + 2 b \rho x (q-1)/T q^2$ and all the off diagonal elements are
equal
$-2 b \rho x/T q^2$, we find a $(q-1)$-fold degenerate eigenvalue $1
+ 2 b \rho x/T q$ and a non-degenerate eigenvalue of $1$ (see the
Appendix for details).
This leads to the energy term of the form
\begin{equation}
\ln \det \left({\widehat I} + {2 \rho x \over T}{\widehat \Delta}
{\widehat B} \right)
	= (q-1) \ln \left(  1 + {2 b \rho x \over q T} \right)
\end{equation}
Note that this term vanishes for the homopolymer ($q=1$) case.
As for the other term of the energy~(\ref{eq:energy}), it reduces to
\begin{equation}
\left<  \vec{p} \left| {\rho \over T} {\widehat B} { \left(
\widehat{I} + {2 \rho x \over T} \widehat{\Delta} {\widehat B}
\right) }^{-1} \right| \vec{p} \right> = \left<  \vec{p} \left| {\rho
\over T} {\widehat B} \right| \vec{p} \right> \ ,
\end{equation}
i.e., to the average second virial term~(\ref{eq:secvirterm}).
This term does not contribute to optimization with respect to $x$.
We find the freezing temperature
\begin{equation}
T_f = {-2 b \rho \over q \Xi(2 s/[q-1])}
\label{eq:TfPotts1}
\end{equation}
where $\Xi(\sigma)$ is given self-consistently by
\begin{equation}
\Xi(\sigma) \ : \ \ \ \ \ \ \ \ \  \sigma = \ln(1 - \Xi) + \Xi/(1 -
\Xi)\simeq\cases{
\Xi ^2 /2 & for $\Xi \ll 1$ \cr
1/(1 - \Xi) & for $\Xi \rightarrow 1$ }
 \ .
\label{eq:XiSelfCon}
\end{equation}
We
see that the freezing temperature decreases with increasing $q$.
Physically, this
corresponds to the fact that in the Potts model, all monomers from
differing
species interact with each other in the same way, so that the part of
the chain without similar monomers is effectively homopolymeric.  As
$q$ increases, these
homopolymer-like regions increase and the freezing temperature
consequently decreases.  When $b$ is negative (positive), we have
physical solutions of $T_f$ for positive (negative) $\Xi$.  We see
from eq.~(\ref{eq:XiSelfCon}) that the nature of the $\Xi$ function
is different positive and negative values: there is a singularity at
$\Xi=1$, whereas $\Xi<0$ is well behaved.  Thus, there is a
fundamental difference between ferromagnetic-like ($b<0$) and
antiferromagnetic-like ($b>0$) interactions in terms of the freezing
behavior.
%

Two simplified asymptotic expressions for $T_f$ can be mentioned,
coming from the two asymptotics of the $\Xi (\sigma)$
function~(\ref{eq:XiSelfCon}):
\begin{equation}
T_f \simeq \cases{
- ( \rho b / \sqrt{s} )  ( \sqrt{q - 1}/ q ) & for effectively
flexible chains, ${2 s/ (q-1)} \ll 1$ \cr
- (2 \rho b / q)[1 +  (q-1)/2 s] & for effectively stiff chains, ${2
s/ (q-1)} \gg 1$
} \ .
\label{eq:TfPotts}
\end{equation}
Recall that the parameter $s= \ln (a^3 / v)$ is related to chain
flexibility \cite{Grosberg}, where $a$ and $v$ are the chain spacer
size and monomer excluded volume, respectively; $s$ is small for
flexible chains, and large for stiff ones. Note, that the regions of
applicability of the two asymptotics in~(\ref{eq:XiSelfCon}) are
controlled by what can be called  the
effective flexibility $\sigma=s/(q-1)$.  Physically, this corresponds
again
to the specific nature of Potts interactions. Indeed, the main
difference between flexible and stiff chains is the number of
neighbors along the chain in the interaction sphere in space around a
given monomer. This number is large for flexible chains and small for
stiff chains. As for Potts interactions, what is relevant is how many
neighbors along the chain {\em attract} a given monomer. This number
is obviously reduced by factor $q-1$, and this explains the
appearance of the effective flexibility $s/(q-1)$.

%
%

\subsubsection {$p$-charge}

In the $p$-charge model \cite{OrlandP,ShakP}, each monomer has a set
of $p$
generalized charges, which can be $s_I^k=\pm 1$.   The Hamiltonian is
defined
to be
\begin{equation}
{\cal H} = \sum_{I,J}^N \delta({\bf r}_I - {\bf r}_J) \sum_{k=1}^p
\chi_k
s_I^k s_J^k
\end{equation}
In the interaction matrix, we define each possible combination of
charges as a
different species.   Thus, there are $q=2^p$ species in the
interaction matrix.
For species number $i$ ($1 \leq i \leq q$), the value of charge $k$
is
given by
$s^k(i) = 2 \left( \left\lfloor i / 2^k \right\rfloor \bmod 2 \right)
- 1$,
where $\left\lfloor \ldots \right\rfloor$ means truncate to the
lowest integer.
Thus, we have an interaction matrix of the form
\begin{equation}
{\widehat B}_{ij} = \sum_k \chi_k
\left[ 2 \left( \left\lfloor {i \over 2^k} \right\rfloor \bmod 2
\right) - 1
\right]
\left[ 2 \left( \left\lfloor {j \over 2^k} \right\rfloor \bmod 2
\right) - 1
\right]
\end{equation}

The ${\widehat \Delta} {\widehat B}$ matrix
has $p$ non-zero eigenvalues $\chi_1, \chi_2, \ldots,\chi_p$
and a $(2^p-p)$-degenerate eigenvalue of 0.  Thus,
\begin{equation}
\ln \det \left({\widehat I} + {2 \rho x \over T}{\widehat \Delta}
{\widehat B} \right)
 =  \sum_{i=1}^p \ln \left( 1 + {2 \chi_i \rho x \over T}  \right)
\label{eq:pchargelog}
\end{equation}
and as in the Potts case,
\begin{equation}
\left<  \vec{p} \left| {\rho \over T} {\widehat B} { \left(
\widehat{I} + {2 \rho x \over T} \widehat{\Delta} {\widehat B}
\right) }^{-1} \right| \vec{p} \right> = \left<  \vec{p} \left| {\rho
\over T} {\widehat B} \right| \vec{p} \right>
\end{equation}
Thus, the freezing temperature is determined by
\begin{equation}
2 s = \sum_{i=1}^p \left[ \ln \left( 1 + {2 \chi_i \rho \over T_f}
\right) -
 {2 \chi_i \rho/T_f \over 1 + 2 \chi_i \rho/T_f} \right]
\end{equation}
For the specific case $\chi_i = \chi$, we have
\begin{equation}
T_f =  - {2 \rho \chi \over \Xi(2 s/p)}  \ ,
\label{eq:Tfpcharge}
\end{equation}
where $\Xi (\sigma)$ function is defined above
by~(\ref{eq:XiSelfCon}).
%
%
As in Potts interactions, the asymmetry of the $\Xi$ function yields
different behavior, depending on the sign of $\chi$.
Unlike the Potts case,
the behavior of the $p$-charge model becomes more heteropolymeric,
i.e. $T_f$ increases, with the addition of more species.

The two asymptotics, for flexible and stiff chains, in $p$-charge
model are
\begin{equation}
T_f \simeq \cases{
- \rho \chi {\left( {2 p / s } \right)}^{1/2} & for effectively
flexible chain, ${s/ p} \ll 1$ \cr
- 2 \rho \chi (1 + p/2 s )	& 	for effectively stiff chain,
${s / p}
\gg 1$
} \ .
\label{eq:Tfpchargelimits}
\end{equation}
Note, that effective flexibility is given by $\sigma =s/p$ for the
$p$-charge model, i.e. it is again reduced by the number of species.

We note that our result~(\ref{eq:Tfpcharge}) reproduces automatically
what is trivially expected for the homopolymer case ($T_f=0$, i.e.,
no freezing, when $p=0$) and also at $p=1$ agrees with our previous
result~(\ref{eq:TfPotts1}) at $q=2$ in the case of two letter Ising
heteropolymer. On the other hand, our equation~(\ref{eq:Tfpcharge}),
or its asymptotics in the first line of eq
(\ref{eq:Tfpchargelimits}), agrees with earlier results of the work
\cite{ShakP} in the opposite extreme of $p \gg 1$, i.e., in the
region of applicability of that work.


\subsection{Reduction Theorems}
\label{sec:reduction}

%
%

There are several cases in which the same physical system can be
depicted in terms of formally different interaction matrices
${\widehat B}$ and/or composition vectors  ${\vec p}$. Clearly, the
expression for the freezing temperature, as well as for any other
real physical quantity, must not depend on any arbitrary choice.

For example, there might be some monomer species which are formally
included in the list, and in the interaction matrix, but they are not
physically presented in the chain, as the corresponding $p$ vanishes,
say, $p_q=0$.  It is easy to check, that in this case
eq~(\ref{eq:finalselfconsist}) is reduced to smaller list of $q-1$
monomer species with $(q-1) \times (q-1)$ interaction matrix.

Another example is when there are duplicate species, say, species
labeled $q$ and $q-1$ are physically identical, i.e. they interact in
identical ways to all other species. Physically, we would expect that
this problem is identical
to the $q-1$ species case, except with the new composition $p_{q-1}'
= p_{q-1} +
p_q$. Even though we skip the proof, eq~(\ref{eq:finalselfconsist})
indeed gives this expected reduction.

These two statements, which we call ``reduction theorems'', are not
only a good check of consistency of our result
{}~(\ref{eq:finalselfconsist}), but they will be also important in
further discussion.


\subsection{Freezing Temperature: General Consideration}

We return to the general analysis of the
equation~(\ref{eq:finalselfconsist}) for the freezing temperature,
and we will show how to implement in the general case both the limits
of stiff and flexible chains, similar to how those cases appear in
the exact solutions for the Potts and $p$-charge models.

We first perform an expansion in powers of $\xi = 2 \rho x/T$. For
example,
\begin{equation}
{\widehat B} {(I + \xi {\widehat \Delta}{\widehat B})}^{-1} =
{\widehat B} - \xi {\widehat B} {\widehat \Delta} {\widehat B} +
\xi^2 {\widehat B} {\widehat \Delta} {\widehat B} {\widehat \Delta}
{\widehat B} + \ldots
\label{eq:BDBDB}
\end{equation}
%
%
%
Note, that any term ${\widehat B} {\left( {\widehat \Delta} {\widehat
B} \right)}^k$,
where $k$ is a positive integer, is independent of $\left< B
\right>$, and therefore
is purely heteropolymeric. The matrix product $\left({\widehat
\Delta} {\widehat B}\right)_{i_1 i_2} \left({\widehat \Delta}
{\widehat B}\right)_{i_2 i_3} \ldots \left({\widehat \Delta}
{\widehat B}\right)_{i_{k-1} i_k}$ can be interpreted as the
propagation of heteropolymeric interactions from monomer species
$i_1$ to $i_2$,  from $i_2$ to $i_3$, etc., up to $i_k$.
%
%
As we suppose from the very beginning that all of the heterogeneity
comes from the second virial coefficient only, so that all higher
order virial terms of the original Hamiltonian are in a sense
homopolymeric, all heteropolymeric interactions are simply pair
collisions of monomers. Each monomer takes part, of course, in a
variety of pair collisions during a very long time, i.e., in
thermodynamic equilibrium. Those collisions are weighted with the
corresponding energies, and they form chains of collisions, described
by ${\widehat B} {\left( {\widehat \Delta} {\widehat B} \right)}^k$
terms. Depending on both the $B_{ij}$ interaction matrix and the
species occurrence probabilities $p_i$, some of those chains might be
more or less favorable than others, and this determines freezing
transition in the system.

To employ the expansion~(\ref{eq:BDBDB}), we first rewrite
(\ref{eq:finalselfconsist}) by noting that $\ln \det {\widehat A} =
{\rm Tr} \ln {\widehat A}$ and $\left< {\vec p} \left| {\widehat A}
\right|  {\vec p} \right> = {\rm Tr}{\widehat P} {\widehat A}$, where
${\widehat P}_{ij}=p_ip_j$:
\begin{equation}
2s = {\rm Tr} \left\{ \ln \left({\widehat I} + \xi {\widehat \Delta}
{\widehat B} \right)
-  \xi {\widehat \Delta} {\widehat B}  { \left[ \widehat{I} +  \xi
{\widehat \Delta} {\widehat B} \right] }^{-1}
+  \xi ^2 {\widehat P} {\widehat B} {\widehat \Delta} {\widehat B}  {
\left[ \widehat{I} +
		\xi {\widehat \Delta} {\widehat B} \right] }^{-2}
\right\}  \ .
\label{eq:finalselfconsist1}
\end{equation}
Now we are in a position to perform the expansion over the powers of
$\xi$, yielding
\begin{equation}
2 s= \sum_{k=2}^{\infty} \xi ^k \left< B ^k \right> _m \ ,
\label{eq:cumulexp}
\end{equation}
where
\begin{equation}
\left< B ^k \right> _m =  {k-1 \over k} {\rm Tr} \left[ \left(-
{\widehat \Delta}  -k {\widehat P} \right) {\widehat B} \left( -
{\widehat \Delta} {\widehat B} \right) ^{(k-1)} \right]  \ .
\label{eq:defcumul}
\end{equation}
The values $\left< B ^k \right> _m$ can be considered as moments of
${\widehat B}$ matrix produced by a given ${\widehat \Delta}$ matrix.
In fact, we can make the substitution $b_{ij} = B_{ij} - \sum_{kl}
p_k p_l B_{kl}$, i.e.
remove the ``homopolymer'' mean from the interaction  matrix, and
the moments can be rewritten exactly with the exchange $B_{ij}
\rightarrow b_{ij}$.
A consequence of this symmetry is that these moments vanish
in the homopolymer case ($b_{ij} = 0$).

We now pass to analysis of two opposite extremes in the
equation~(\ref{eq:finalselfconsist1}).

\subsubsection{Freezing Temperature: Stiff Chain Limit}

%
%
As we are instructed by the examples of Potts and $p$-charge models,
what is important in high $s$ limit is the singularity of the right
hand side of~(\ref{eq:finalselfconsist1}). This is obviously governed
by high $k$ terms of power series, which are basically related to
$\left(- {\widehat \Delta} {\widehat B} \right)^{k} $. This is
reminiscent of the standard problems of $1D$ statistical physics,
such as the $1D$ Ising model, the ideal polymer, or other Markovian
processes, where $\left(- {\widehat \Delta} {\widehat B} \right)$
plays the role of the transfer matrix. It is well known that highest
eigenvalue of the transfer matrix is only relevant in $k \rightarrow
\infty$ limit (``ground state dominance principle''). In this limit,
$\xi \simeq 1 / \lambda_{\rm max}$, where $\lambda_{\rm max}$ is
highest eigenvalue of  $\left(- {\widehat \Delta} {\widehat B}
\right)$ matrix, and thus
\begin{equation}
T_f \simeq 2 \rho \lambda_{\rm max}  \ .
\end{equation}
To find the next terms in asymptotic formula for $T_f$, we note that
the most divergent term in eq ~(\ref{eq:cumulexp}) comes form the
last term in ~(\ref{eq:finalselfconsist1}) and is due to $k {\widehat
P}$ term in ~(\ref{eq:defcumul}), it diverges as $(1- \xi
\lambda_{max})^{-2}$. We know, however, that this term vanishes for
both Potts and $p$-charge models. Moreover, we can show, that it
vanishes also for many other models with some regularities, producing
cancellation of correlations and anti-correlations between matrix
elements of ${\widehat B}$. For this reason, we keep next to the
highest singularity, thus obtaining
\begin{equation}
2s \simeq {c \over (1 - \xi \lambda_{max})^2 }+{1 \over (1 - \xi
\lambda_{max})} \ ; \ \
c = \left< {\vec p} \left| \right. {\vec \psi} \right> \left< {\vec
\psi} \left| -{\widehat B} / \lambda \right| {\vec p} \right>
\end{equation}
where $\lambda$ and $\left| {\vec \psi} \right>$ are the eigenvalue
and the corresponding eigenvector of the $\left(- {\widehat
\Delta}{\widehat B} \right)$ operator.
This gives finally
\begin{equation}
T_f \simeq 2 \rho \lambda_{\rm max} \left[1 + {1 +\sqrt{1 +4cs} \over
2s} \right] \simeq \cases{
2 \rho \lambda_{\rm max} \left[1 +\sqrt{c/s}  \right] & for $cs \gg 1
\ \ \ \ (c \neq 0) $ \cr
2 \rho \lambda_{\rm max} \left[1 + 1/s \right] & for $cs \ll 1  \ \ \
\ (c = 0)$ }
  \ .
\end{equation}

Note that $\lambda_{max}$, as an eigenvalue, depends strongly on the
arrangement of matrix
elements.  Therefore freezing transition for stiff chains
is very dependent on the pattern of interactions, not only on their
overall heterogeneity. This has clear physical meaning. In case of
stiff chains, real
monomers represent the physical units of interaction. In other words,
quasi monomers
almost coincide with monomers. In terms of propagation, or chains of
collisions (see
above), it is clear that highest eigenvalue of $\left(- {\widehat
\Delta} {\widehat B} \right)$
matrix corresponds to the lowest (because of the sign) energy of
interaction, while the
corresponding eigenvector, in terms of the obvious quantum mechanical
analogy, is the
linear combination of monomers which realizes this lowest energy and
thus controls the
freezing temperature.

\subsubsection{Freezing Temperature: Flexible Chain Limit}

The examination of the small $s$ case may be on the first glance
questionable, as our approach is entirely mean-field in nature and,
therefore, it might be applicable for large enough $s$ only. We have
seen, however, in the examples of Potts and $p$-charge models, that
the applicability of the flexible chain limit is controlled by the
{\em effective} flexibility, which is considerably smaller than $s$
itself. We therefore consider formally the small $s$ limit, leaving
the analysis of applicability for each particular case.

In small $s$ limit, only the first term with $k=2$ is relevant in the
series~(\ref{eq:cumulexp}). Omitting all higher order terms,  we
obtain the remarkably simple result
\begin{equation}
T_f =     {2  \rho \over \sqrt{s}}  \left<  {\widehat B}^2
\right>_c^{1/2}
\label{eq:Tfvar}
\end{equation}
where the second cumulant (variance) is defined as $\left<  {\widehat
B}^2 \right>_c \equiv \left< \left[ {\widehat B} - \left<  {\widehat
B} \right> \right]^2 \right>$ and
matrix averages are defined by
\begin{equation}
\left< {\widehat B} \right> \equiv  \sum_{ij} p_i p_j  B_{ij}  \ .
\end{equation}
Unlike the stiff chain limit, in the flexible chain case at hand, the
freezing transition is controlled mainly by overall heterogeneity of
interaction energies $B_{ij}$. Thus, if one started with an
interaction matrix with independent elements and
shuffled the matrix elements (even though it is hard to think of real
physical experiment of this kind), this transformation
would not change the freezing temperature for flexible chains. This
is qualitatively a very natural result, as the nature of flexible
chains is such that for any given monomer, many of the neighbors in
space are neighbors along the chain. In other words, the interaction
units are quasi monomers, which are substantially different from the
monomers and represent clouds of monomers, where the individuality of
each monomer species (with different patterns of energetical
preferences to other species) is lost.

In the case of the Potts and $p$-charge models, the variance of the
interaction matrix yields the flexible chain limits for both the
Potts~(\ref{eq:TfPotts}) and $p$-charge~(\ref{eq:Tfpcharge}) models.
 Thus, the solution~(\ref{eq:Tfvar}) for $T_f$ in this limit is
remarkably simple and powerful.  To demonstrate this, we show some
particular examples.

\subsection{Independent Interaction Model}
In the Independent Interaction Model, all $B_{IJ}$ are taken
independently from Gaussian distribution
\begin{equation}
P\left( B_{IJ} \right) =  {\left( { {\tilde B}^2 \over 2 \pi}
\right)}^{1/2}
\exp \left[ - { {\left( B_{IJ} - B_0 \right) }^2 \over {\tilde B}^2 }
\right]
\label{eq:gaussprob}
\end{equation}
(recall that capital $I$ and $J$ are related to monomer numbers along
the chain and not to species). From the physical point of view, the
independence of, say, $B_{IJ}$ and $B_{JK}$ can be realized if and
only if the total number of different species is very large, i.e., in
the $q \rightarrow \infty$ limit. The effective stiffness in this
limit is small, and we have to use the expression~(\ref{eq:Tfvar})
for the freezing temperature. Therefore,  $T_f = 2 \rho {\tilde
B}/\sqrt{s}$. This indeed coincides with original result of the work
\cite{IndepenInt}.

\subsection{Random Sequences of Real Amino Acids}

It is of special interest to examine the freezing transition for
polymers comprised of real amino acids, i.e., of constituents of real
proteins. This can be done using the matrix of interaction energies
derived for amino acids by Miyazawa and Jernigan \cite{Jernigan}. We
are in a position to examine the freezing transition for random
sequences (even though real protein sequences might not be random
\cite{Ptitsyn,ProtCor}). In the work \cite{Jernigan}, the interaction
matrix was given in the form $U_{ij}/T_{MJ}$, where $U_{ij}$ is the
interaction energy and $T_{MJ}$ is a temperature not formally defined
in \cite{Jernigan}.   In some rough approximation, we identify the MJ
matrix with our $\rho {\widehat B}$. To avoid rewriting of the eq
{}~(\ref{eq:finalselfconsist}), we substitute the MJ matrix
into~(\ref{eq:finalselfconsist}) instead of ${\widehat B}$, meaning
that now $\xi = 2 T_{MJ}/T_f$. We assume also equal composition $p_i
=1/q =1/20$. We can then numerically
calculate the $\xi$ vs $s$ dependence.  The result is shown in
Figure~1.  Note the
qualitative similarity of the graph of $\xi$ vs $s$ for the MJ matrix
and $\Xi$ vs $s$ given by~(\ref{eq:XiSelfCon}).

Given the realistic value of $s \approx 1.4$ ($v/a^3 \approx 0.25$)
for polypeptide chain, we obtain from the Figure~1 the estimate $\xi
\approx 1.6$, or $T_f \approx 1.25 T_{MJ}$. By taking more realistic
uneven composition, we arrive at  $\xi \approx 1.75$, or $T_f \approx
1.14 T_{MJ}$.  Note that for real amino acids system the relevant
solution is generally in the high flexibility regime.

To understand these results, recall the way that the MJ matrix was
derived in \cite{Jernigan}. The protein $3D$ structures data bank was
employed such that if there were ${\cal M}_{ij}$ contacts between
amino acids labelled as $i$ and $j$ in the data bank, and the total
number of contacts was ${\cal M}$, then the ratio ${\cal
M}_{ij}/{\cal M}$ was interpreted as a probability governed by some
effective Boltzmann distribution  ${\cal M}_{ij}/{\cal M}= \exp
\left[ - U_{ij}/T_{MJ} \right]$, thus yielding the MJ matrix of
energies, $U_{ij}$. In the later work \cite{Gutin}, it was shown that
the ratio ${\cal M}_{ij}/{\cal M}$ obeys indeed Boltzmann type
formula if proteins do match the random energy model, and then the
parameter of distribution, $T_{MJ}$, is nothing but the freezing
temperature, $T_f$ \cite{GutinThank}.  From that logic, we expect
thus $T_f = T_{MJ}$. Our result is slightly higher.  We conclude
thus, that there is a reasonable agreement between the works
\cite{Jernigan}, \cite{Gutin}, and our results.

%
%

\section{Conclusion}

Starting from a sequence-model Hamiltonian in which interactions
between {\em species} of monomers is expressed in terms of
some arbitrary symmetric matrix ${\widehat B}$, we have derived a
formalism with
which to examine the freezing transition of random heteropolymers.
As monomer species interactions are given by some matrix, this
formulation is the most general form, assuming that interactions are
short range and that heteropolymeric contributions come primarily
from two-body interactions.

First, we have related the freezing temperature to the interaction
matrix
self-consistently.  This self-consistent equation can be solved
exactly
for certain specific systems.  For example, models such as the
$q$-Potts and $p$-charge models are important as they describe
interesting physical cases, but with only a minimal amount of
complexity in their solutions.  It is especially interesting that
these two
simple models have radically different freezing behavior with respect
to the number
of species.  Clearly simply adding new and different monomer species
does
not necessarily enhance the freezing transition.

Taking another approach, we can trade the accuracy of an exact result
for the generality of the assumption of only some arbitrary symmetric
interaction matrix.  To this end, we solved the exact self-consistent
equation perturbatively.  Due to the nature of the $\Xi$ function,
there are two regimes of interest: small $s$ (high effective
flexibility) where $\Xi \rightarrow 0$ and $s \rightarrow \infty$,
where $\Xi$ approaches  a singularity in the self-consistent
formulation.  Expanding at these two limits, we found
$$
T_f \simeq \cases{
(2  \rho / \sqrt{s})  \left< {\widehat B}^2  \right>_c^{1/2}  & for
small $s$ \cr
\rho \lambda_{\rm max} & for large $s$
}
$$
where $\lambda_{\rm max}$ is the largest eigenvalue of the
$-{\widehat \Delta} {\widehat B}$ matrix.

The equation above quantitatively details certain descriptions of
what one could qualitatively call the ``heteropolymeric character''
of the interaction matrix ${\widehat B}$ and the species composition
${\vec p}$.  Specifically, for flexible chains, one would expect that
the physical unit of interactions, or quasi monomers, consist of
several monomers.  The variance of the interaction matrix gives, in a
sense, the heteropolymeric width of interactions.  If these
interaction energies are ordered in the interaction matrix, however,
the correlations between monomer species interactions reduces the
heteropolymeric nature of the system, and thus reduces the freezing
temperature.

In the limit of stiff chains, quasi monomers generally consist of
individual monomers.  Thus, the specific nature of interactions are
of paramount importance.  In this limit, one can imagine the
interactions in space (i.e. not necessarily along the chain) as
interactions propagating through the pairwise interactions of monomer
species.  This chain of interactions, in the stiff polymer limit,
becomes very long and thus the system shares characteristics with
other one-dimensional systems, such as the 1D Ising model.
Specifically, here the freezing temperature is proportional to the
largest eigenvalue, which dominates in the long interaction chain
limit, of the transfer matrix $-{\widehat \Delta} {\widehat B}$.

To conclude, we comment on the applicability of this theory.  Three
points are to be mentioned here. First, since we truncate the series
{}~(\ref{eq:source_term_series}) to ${\cal O}(\phi^2)$, we cannot
comment on the nature of any phase transitions in the average value
of $\phi$, such as the microphase segregation seen in two letter
ferromagnetic interaction matrices \cite{Sfatos}.  Second, all
calculations have been mean field and therefore effects due to the
fluctuations in the order parameter $Q_{\alpha \beta}$ have been
neglected.  At present, these effects have been examined in one-loop
approximation only \cite{Igor,Zhenya}.  For further details, we refer
the reader to the discussions in \cite{Sfatos,q2Design}.  Even though
the issue remains open, to the best of our understanding at this
moment, mean field $\phi^2$ theory is definitely applicable in the
vicinity of freezing transition, at least when the number of monomer
species is more than two. Third, and last, we have ignored the
possibility of liquid crystalline ordering in the stiff chain regime.
This issue has not been examined at all, we can only guess
qualitatively that orientational ordering should be suppressed in
strongly non-uniform heteropolymer system compared to a homopolymer
of comparable stiffness.

Thus, for models with ``heteropolymeric character,'' i.e., the
interaction matrix and probability distribution cannot be reduced to
that of a homopolymer, our theory predicts a freezing transition.
Our formalism facilitates the calculation of specific models of
interactions, but perhaps most importantly, the direct relationship
between the interaction matrix and the freezing transition is
demonstrated.

\bigskip

\centerline{\bf ACKNOWLEDGEMENTS}

The work was supported by NSF (DMR 90-22933)  and NEDO of Japan.
VSP acknowledges the support of an NSF Fellowship.
AYG acknowledges the support of Kao Fellowship.

\newpage

%
%

\appendix

\section{Proof of equation~(20)}

{\bf 1.} Consider first the auxiliary problem of some $x \times x$
matrix ${\widehat q}^{(x)}$ with diagonal elements $\tilde{q}$
and off diagonal
elements $q$. This matrix has a $(x-1)$-fold degenerate eigenvalue
$\lambda=\tilde{q}-q$, corresponding to the eigenvectors $ \left
(\matrix{
1 & -1 & 0 & 0 & \ldots & 0 \cr}\right )$, $\left (\matrix{
1 & 0 & -1 & 0 & \ldots & 0 \cr}\right )$, $\ldots$, $ \left
(\matrix{
1 & 0 & \ldots & 0 & -1 & 0 & \ldots & 0 \cr}\right )$,  $\ldots$ ,
$\left (\matrix{
1 & 0 & 0 & \ldots & 0 & -1 \cr}\right )$, and a non-degenerate
eigenvalue of
$\lambda=\tilde{q}+(x-1)q$, corresponding to the eigenvector $\left
(\matrix{
1 & 1 & 1& \ldots & 1 \cr
}\right )$. Of course, there are other ways of choosing eigenvectors,
in particular, we can built up orthonormal basis by choosing
\begin{equation}
 {\cal R}_{\alpha \beta} = {1 \over \sqrt{x}} \exp \left[ {2 \pi i
\over x} (\alpha - 1) (\beta - 1) \right] \ ; \ \ \ \ \ \ \ \ \ \ \ \
\ \ \ 1 \leq \alpha , \beta \leq x \ .
\label{eq:orthonormbasis}
\end{equation}
Here $\alpha$ numerates eigenvectors, while $\beta$ numerates
components of the given eigenvector (or vice versa).  We can
interpret ${\widehat {\cal R}}^{(x)}= {\cal R}_{\alpha \beta}$ as the
unitary operator transforming ${\widehat q}^{(x)}$ to diagonal form,
${\widehat {\cal R}}{\widehat q}{\widehat {\cal R}}^{-1}={\widehat
\lambda}^{(x)} \equiv \lambda _{\alpha} \delta_{\alpha \beta}$, with
the eigenvalues $\lambda _{\alpha}$ given above.
%
%
\footnote{For completeness, we write also the inverse of ${\widehat
q}^{(x)}$: it has diagonal elements $(\tilde{q}-q)^{-1} -q\{
(\tilde{q}-q) [\tilde{q}+(x-1) q]\}^{-1}$ and off diagonal elements
$-q\{ (\tilde{q}-q) [\tilde{q}+(x-1) q]\}^{-1}$.}
%
%
We will be particularly interested in the case $q=\tilde{q}=1$. In
this case, the non-degenerate eigenvalue is $\lambda =1$, while all
the others are zero.

{\bf 2.} Consider now some general properties of the ``direct
product'' operation for matrices. We repeat the definition:
${\widehat A}^{(r)} \otimes {\widehat B}^{(s)}$ is $rs \times rs$,
built up by substitution of $s\times s$ block  $A_{uv} {\widehat
B}^{(s)}$ instead of each matrix element of  ${\widehat A}^{(r)}$.
\begin{enumerate}
\item By matrix row and column operations, it is easy to show that
the rule is commutative, i.e. \begin{equation}
{\widehat A}^{(r)} \otimes {\widehat B}^{(s)}={\widehat B}^{(s)}
\otimes {\widehat A}^{(r)} \ .
\end{equation}

\item {\em Block matrix multiplication rule}:  it is well known that
the operation of block matrix multiplication is carried out in the
same
scheme as normal matrix multiplication, except the multiplication
of elements is replaced by the matrix multiplication of blocks. This
can be written as
\begin{equation}
\left( {\widehat A}^{(r)} \otimes {\widehat B}^{(s)} \right) \cdot
\left(
 {\widehat A}^{\prime (r)}
 \otimes
{\widehat B}^{\prime (s)} \right) = \left( {\widehat A}^{(r)}
{\widehat A}^{\prime (r)} \right) \otimes \left( {\widehat B}^{(s)}
{\widehat B}^{\prime (s)} \right) \ .
\end{equation}

\item Commutation of $ {\widehat A}^{(r)} \otimes {\widehat B}^{(s)}
$ and ${\widehat A}^{\prime (r)} \otimes {\widehat B}^{\prime (s)}$
depends on commutation of {\em both} pairs $ {\widehat A}^{(r)}
\&{\widehat A}^{\prime (r)}$ and $ {\widehat B}^{(s)} \& {\widehat
B}^{\prime (s)}$ (this directly follows from previous.)

\item The determinant of a block diagonal matrix equals to the
product of determinants of
the diagonal blocks. In particular,
\begin{equation}
\det \left(  {\widehat A}^{(r)} \otimes {\widehat I}^{(s)} \right) =
\left( \det   {\widehat A}^{(r)}  \right)^s
\end{equation}

\item The definition of direct product can be trivially generalized
for non-square matrices and, in particular, for vectors
%
%
\footnote{${\widehat A}^{(r \times r^{\prime})} \otimes {\widehat
B}^{(s \times s^{\prime})}$ is generally the matrix $rs \times
r^{\prime} s^{\prime}$}
%
%
. For example, $ \left| \vec{\rho}^{(nq)}\right>=
\vec{\rho}^{(n)}\otimes \vec{p}^{(q)}$.

\item {\em Matrix operation with a vector}:
\begin{equation}
 {\widehat A}^{(r)} \otimes {\widehat B}^{(s)}  \left| \vec{a}^{(r)}
\otimes \vec{b}^{(s)}\right> = {\widehat A}^{(r)} \left|
\vec{a}^{(r)} \right> \otimes {\widehat B}^{(s)}  \left|
\vec{b}^{(s)}\right>
\end{equation}

\item {\em Scalar product of vectors}:
\begin{equation}
\left< {\vec a}^{(r)} \otimes {\vec b}^{(s)} \right| \left.
\vec{a}^{\prime (r)} \otimes \vec{b}^{\prime (s)} \right> = \left<
{\vec a}^{(r)}  \right| \left. \vec{a}^{\prime (r)} \right> \left<
{\vec b}^{(s)} \right| \left.  \vec{b}^{\prime (s)} \right>
\end{equation}

\end{enumerate}

The proof of all the above mentioned properties is straightforward.

{\bf 3.} Let us return now to the expression of energy
{}~(\ref{eq:energy1}). We have to address the matrix $\left[
\widehat{I}^{(qn)}+{2 \rho  \over T}
 {\widehat q}^{(n)} \otimes {\widehat \Delta}^{(q)}
{\widehat B}^{(q)} \right]$. We know (or we assume) that ${\widehat
q}^{(n)}$ is comprised of $n/x$ ${\widehat q}^{(x)}$ blocks along the
diagonal, with $\tilde{q}=q=1$, that is ${\widehat q}^{(n)}={\widehat
I}^{(n/x)} \otimes {\widehat q}^{(x)}$. First, this form of
${\widehat q}^{(n)}$ matrix allows us to factor the matrix of our
interest:
\begin{equation}
\left[ \widehat{I}^{(qn)}+{2 \rho  \over T}
 {\widehat q}^{(n)} \otimes {\widehat \Delta}^{(q)}
{\widehat B}^{(q)} \right] = \widehat{I}^{(n/x)} \otimes \left[
\widehat{I}^{(qx)}+{2 \rho  \over T}
 {\widehat q}^{(x)} \otimes {\widehat \Delta}^{(q)}
{\widehat B}^{(q)} \right] \ .
\label{eq:factor}
\end{equation}
This means physically that replicas of different groups are not
coupled, they  do not interact to each other.

The remainder (in the square brackets in the right hand side
of~(\ref{eq:factor})) can be diagonalized via the rotation operator
${\widehat {\cal R}}^{(xq)} = {\widehat {\cal R}}^{(x)} \otimes
{\widehat I}^{(q)} $. Indeed, using properties 2 and 3 above, we
have:
\begin{equation}
\left( {\widehat {\cal R}}^{(xq)} \right)^{-1} \left[
\widehat{I}^{(xq)}+ {2 \rho \over T}  {\widehat q}^{(x)} \otimes
{\widehat \Delta}^{(q)} {\widehat
B}^{(q)}
\right]  {\widehat {\cal R}}^{(xq)}
 = \widehat{I}^{(xq)}  +  {2 \rho \over T} {\widehat \lambda}^{(x)}
\otimes {\widehat \Delta}^{(q)}
{\widehat B}^{(q)} \ .
\label{eq:rotation}
\end{equation}
Recall that there is only one non-zero $\lambda$, and therefore the
last matrix has one $q \times q$ block $\left(2 \rho/T\right)
{\widehat \Delta}^{(q)}
{\widehat B}^{(q)}$ in the upper-left corner, it has 1 down this
block on the main diagonal, and all other matrix elements are 0.

We are now in a position to simplify the first term of
energy~(\ref{eq:energy1}). First, we apply the rule~4 to this energy
term, then we note that determinant does not change upon
rotation~(\ref{eq:rotation}), while the determinant of the right hand
side of~(\ref{eq:rotation}) is trivially computed, yielding
\begin{equation}
\ln \det \left[ \widehat{I}^{(qn)}+ {2 \rho \over T} {\widehat
q}^{(n)} \otimes {\widehat
\Delta}^{(q)} {\widehat B}^{(q)} \right] =
{n \over x} \ln  \det \left[ \widehat{I}^{(q)} + {2 \rho x \over T}
{\widehat \Delta}^{(q)} {\widehat B}^{(q)}
 \right]
\label{eq:flucreduc}
\end{equation}

As for the second term in~(\ref{eq:energy1}), we first apply the
rule~7 to get
\begin{eqnarray}
& & \left< {\vec \rho}^{(n q)} \left| {1 \over T} {\widehat B}^{(q)}
\otimes {\widehat I}^{(n)}
\left[ \widehat{I}^{(qn)}+{ 2 \rho \over T}
{\widehat q}^{(n)} \otimes {\widehat \Delta}^{(q)} {\widehat B}^{(q)}
 \right]^{-1} \right| {\vec \rho}^{(n q)} \right> = \nonumber \\
& = &  {n \over x}  \left< {\vec \rho}^{(x q)} \left| \left(  {1
\over T} {\widehat B}^{(q)} \otimes {\widehat I}^{(x)} \right) \left[
\widehat{I}^{(qx)}  +  {2 \rho \over T} {\widehat q}^{(x)} \otimes
{\widehat \Delta}^{(q)}
{\widehat B}^{(q)} \right] ^{-1} \right| {\vec \rho}^{(x q)} \right>
\ .
\end{eqnarray}
We then use the rotation~(\ref{eq:rotation}) and note that $
{\widehat B}^{(q)} \otimes {\widehat I}^{(x)}$ and $ {\widehat {\cal
R}}^{(qx)}$ do commute to each other due to the rule~3. This yields
the form
\begin{equation}
{n \over x}  \left< {\vec \rho}^{(x q)} \left| {\left( {\widehat
{\cal R}}^{(x)} \otimes {\widehat I}^{(q)} \right)}^{-1}  \left(  {1
\over T} {\widehat B}^{(q)} \otimes {\widehat I}^{(x)} \right) \left[
\widehat{I}^{(qx)}  +  {2 \rho \over T} {\widehat \lambda}^{(x)}
\otimes {\widehat \Delta}^{(q)}
{\widehat B}^{(q)} \right]^{-1} \left( {\widehat {\cal R}}^{(x)}
\otimes {\widehat I}^{(q)} \right) \right| {\vec \rho}^{(x q)}
\right> \ .
\end{equation}

We consider, therefore, the rotation of density vector $\left| {\vec
\rho}^{(x q)} \right>$. First, we note that $ {\vec \rho}^{(x q)}=
{\vec \rho}^{(x )}\otimes {\vec p}^{(q)}$.  Second,  the density, as
the physical quantity, is the same for all replicas and does not
depend on replica indices. To write it formally, let us define two
$x$-dimensional vectors ${\vec i}^{(x)}=\left (\matrix{ 1 & 1 & 1 &
\ldots &1} \right)$  and ${\vec j}^{(x)}=\left (\matrix{ 1 & 0 & 0 &
\ldots &0} \right)$. Then we see by direct implementation of
formula~({eq:orthonormbasis}) $ {\widehat {\cal R}}^{(x)} \left|
{\vec i}^{(x)} \right>= \sqrt{x} {\vec j}^{(x)}$. On the other hand,
$\vec{\rho}^{(x)}=\rho  {\vec i}^{(x)}$. Therefore, according to the
rule~5, we have ${\widehat {\cal R}}^{(xq)} \left| {\vec \rho}^{(x
q)} \right> = \rho \sqrt{x} {\vec j}^{(x)} \otimes {\vec p}^{(q)}$.
This yields the energy term in the form
\begin{equation}
{n \over x} \left<\rho \sqrt{x} {\vec j}^{(x)} \otimes {\vec p}^{(q)}
\right|  \left(  {1 \over T} {\widehat B}^{(q)} \otimes {\widehat
I}^{(x)} \right) \left[ \widehat{I}^{(qx)}  +  {2 \rho \over T}
{\widehat \lambda}^{(x)} \otimes {\widehat \Delta}^{(q)}
{\widehat B}^{(q)} \right]^{-1} \left| \rho \sqrt{x} {\vec j}^{(x)}
\otimes {\vec p}^{(q)} \right>
\end{equation}
As ${\vec j}^{(x)}$ has only one non-zero-component, and ${\widehat
\lambda}^{(x)}$ has also only one non-zero matrix element,
corresponding to the same direction in vector $x$-dimensional space,
we have
\begin{equation}
\left[ \widehat{I}^{(qx)}  +  {2 \rho \over T} {\widehat
\lambda}^{(x)} \otimes {\widehat \Delta}^{(q)}
{\widehat B}^{(q)} \right]^{-1} \left|  {\vec j}^{(x)} \otimes {\vec
p}^{(q)} \right> =  {\vec j}^{(x)}\otimes  \left[ {\widehat I}^{(q)}+
{\widehat \Delta}^{(q)} {\widehat B}^{(q)} \right]^{-1} {\vec
p}^{(q)} + \left(  {\vec i}^{(x)}- {\vec j}^{(x)} \right) \otimes
{\vec p}^{(q)} \ .
\end{equation}
The last step is to implement the scalar product rule 7, yielding
\begin{equation}
{n \over x} \rho^2 x \left< {\vec p}^{(q)} \right|  \left(  {1 \over
T} {\widehat B}^{(q)}  \right) \left[ \widehat{I}^{(q)}  +  {2 \rho
\over T}  {\widehat \Delta}^{(q)}
{\widehat B}^{(q)} \right]^{-1} \left| {\vec p}^{(q)} \right> \ .
\label{eq:mftermreduc}
\end{equation}
Combining~(\ref{eq:flucreduc}) with (\ref{eq:mftermreduc}), we arrive
at (\ref{eq:energy}).

\newpage


\centerline{\bf FIGURE CAPTIONS}

\noindent \underline{Figure 1}:

\noindent a)  Plot of the inverse reduced freezing temperature
($\Xi$) vs the effective flexibility ($\sigma$), with the inset of
the graph showing the detail of the small $s$ vs small $\Xi$ regime.
The important characteristics of this
function is that it is described by $\Xi^2$ for small $\Xi$ and the
existence of a singularity at $\Xi=1$.  The solid line denotes the
exact solution, the unevenly dashed line denotes the stiff chain
expansion, and the evenly dashed line denotes the flexible chain
expansion.

\noindent b)  For the Miyazawa and Jernigan matrix of amino acid
interactions, we plot the flexibility ($s$) vs the reduced inverse
freezing temperature ($\xi$), with the inset of the graph showing the
detail of the small $s$ vs small $\xi$ regime.  Qualitatively, this
curve is similar to $\Xi(\sigma)$. Further note, however, that any
physical polymer will be described by the small $\xi$ regime.  The
dashing of curves denotes the same approximations as in part (a).


\begin{thebibliography}
\bibitem{GutinThank} We are indebted to A.Gutin for the discussion of
this point.
\end{thebibliography}
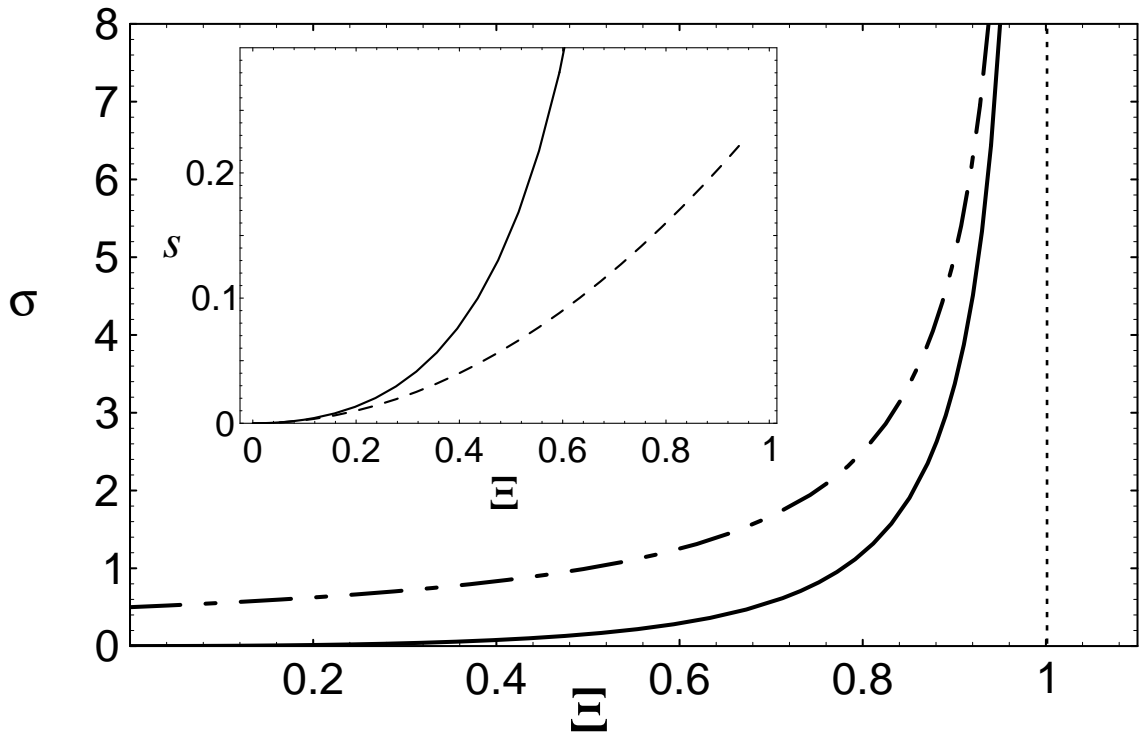
\end{document}